\documentclass{optica-article}

\journal{oe}

\articletype{Research Article}

\usepackage{lineno}
\usepackage{comment}
\usepackage{amsmath,amssymb}
\usepackage{graphicx}
\usepackage[detect-weight, per-mode=symbol, separate-uncertainty=true]{siunitx}

\begin{document}

\title{On-target delivery of intense ultrafast laser pulses through hollow-core anti-resonant fibers}

\author{Athanasios Lekosiotis\authormark{1,*}, Federico Belli\authormark{1}, Christian Brahms\authormark{1}, Mohammed Sabbah\authormark{1}, Hesham Sakr\authormark{2}, Ian A. Davidson\authormark{2}, Francesco Poletti\authormark{2}, and John C. Travers\authormark{1}}

\address{\authormark{1}School of Engineering and Physical Sciences, Heriot-Watt University, Edinburgh, EH14 4AS, UK\\
\authormark{2}Optoelectronics Research Centre, University of Southampton, Southampton, SO17 1BJ, UK}

\email{\authormark{*}a.lekosiotis@hw.ac.uk} 



\begin{abstract}
We report the flexible on-target delivery of 800~nm wavelength, 5~GW peak power, 40~fs duration laser pulses through an evacuated and tightly coiled 10~m long hollow-core nested anti-resonant fiber by positively chirping the input pulses to compensate for the anomalous dispersion of the fiber. Near-transform-limited output pulses with high beam quality and a guided peak intensity of $3~\text{PW/}\text{cm}^2$ were achieved by suppressing plasma effects in the residual gas by pre-pumping the fiber with laser pulses after evacuation. This appears to cause a long-term removal of molecules from the fiber core. Identifying the fluence at the fiber core-wall interface as the damage origin, we scaled the coupled energy to 2.1~mJ using a short piece of larger-core fiber to obtain 20~GW at the fiber output. This scheme can pave the way towards the integration of anti-resonant fibers in mJ-level nonlinear optical experiments and laser-source development.
\end{abstract}

\section*{Introduction}
Hollow-core micro-structured fibers have proven to be an excellent platform for high-power laser pulse delivery due to the low light-glass interaction they offer while maintaining low loss~\cite{Shephard:04,debord_multi-meter_2014, emaury_efficient_2014}. Although their optical damage threshold is commonly determined by the input pulse energy density (fluence)~\cite{smith_optical_2009} when operating in the picosecond~\cite{michieletto_hollow-core_2016} or few-hundred-femtosecond regime~\cite{debord_multi-meter_2014}, it has not been widely studied for laser pulses with durations of a few tens of femtoseconds. In this regime, very high intensity can be present in the fiber core even for moderate pulse energies, possibly exceeding the ionisation threshold intensity of helium of $0.5~\text{PW/}\text{cm}^2$~\cite{boreham_measurement_1981}, which is the highest among noble gases. As a consequence, high-field applications have been limited to pulse energies below \SI{100}{\micro\joule}~\cite{balciunas_strong-field_2015} and routes towards ultrafast mJ-level pulse delivery have remained unexplored. While simple hollow capillaries have enabled the transmission of extreme intensities, exceeding $100~\text{PW/}\text{cm}^2$~\cite{Cros2002,Cros2004}, they cannot be used as a flexible waveguide for pulse delivery due to prohibitively high bend loss and mode mixing~\cite{Cros2002,marcatili_hollow_1964}. In contrast, nested negative-curvature anti-resonant hollow fibers~\cite{poletti_nested_2014, sakr_hollow_2020} exhibit exceptionally high transmission (with attenuation as low as 0.5~dB/km at 850~nm) and low bend loss, and so are an optimal choice for pushing the boundaries of flexible ultrashort laser pulse delivery over long lengths~\cite{newkirk_high_2021, mulvad_kilowatt-average-power_2022}.

In this work we first investigated the limits of a 10~m long nested anti-resonant hollow fiber, coiled tightly with a 6~cm radius, in terms of both maximum fluence and guided intensity, and subsequently achieved the efficient delivery of high-power compressed pulses to a target after the fiber by pre-compensating for the anomalous dispersion of the fiber. At the fiber output, we obtained 40~fs compressed pulses with \qty{200}{\micro\joule} pulse energy, 5~GW peak power and excellent beam quality. Furthermore, we observed a suppression of plasma effects after pre-pumping the fiber at high intensity following evacuation. This appears to cause a surprising long-term removal of molecules from the fiber core, improving output pulse quality and providing high transmission. Such results can pave the way for efficient delivery of high-intensity pulses and experiments on a target at distant locations.

Despite the high guided peak intensity of $3~\text{PW/}\text{cm}^2$, we identified the fluence at the inner interface of the core-wall as the primary origin of damage, similar to hollow capillaries~\cite{Cros2002,Cros2004}. Thus, mJ-level pulse delivery can be accessed by scaling the fiber core size. We verified this experimentally by coupling \SI{1.8}{\milli\joule} of pulse energy into a short piece of larger-core anti-resonant fiber, corresponding to 20~GW peak power at the fiber output, in accordance with previous results and the fluence-induced damage threshold predicted for silica glass.

\section{Experimental setup}

\begin{figure}[ht]
\centering
\includegraphics[width=4.25in]{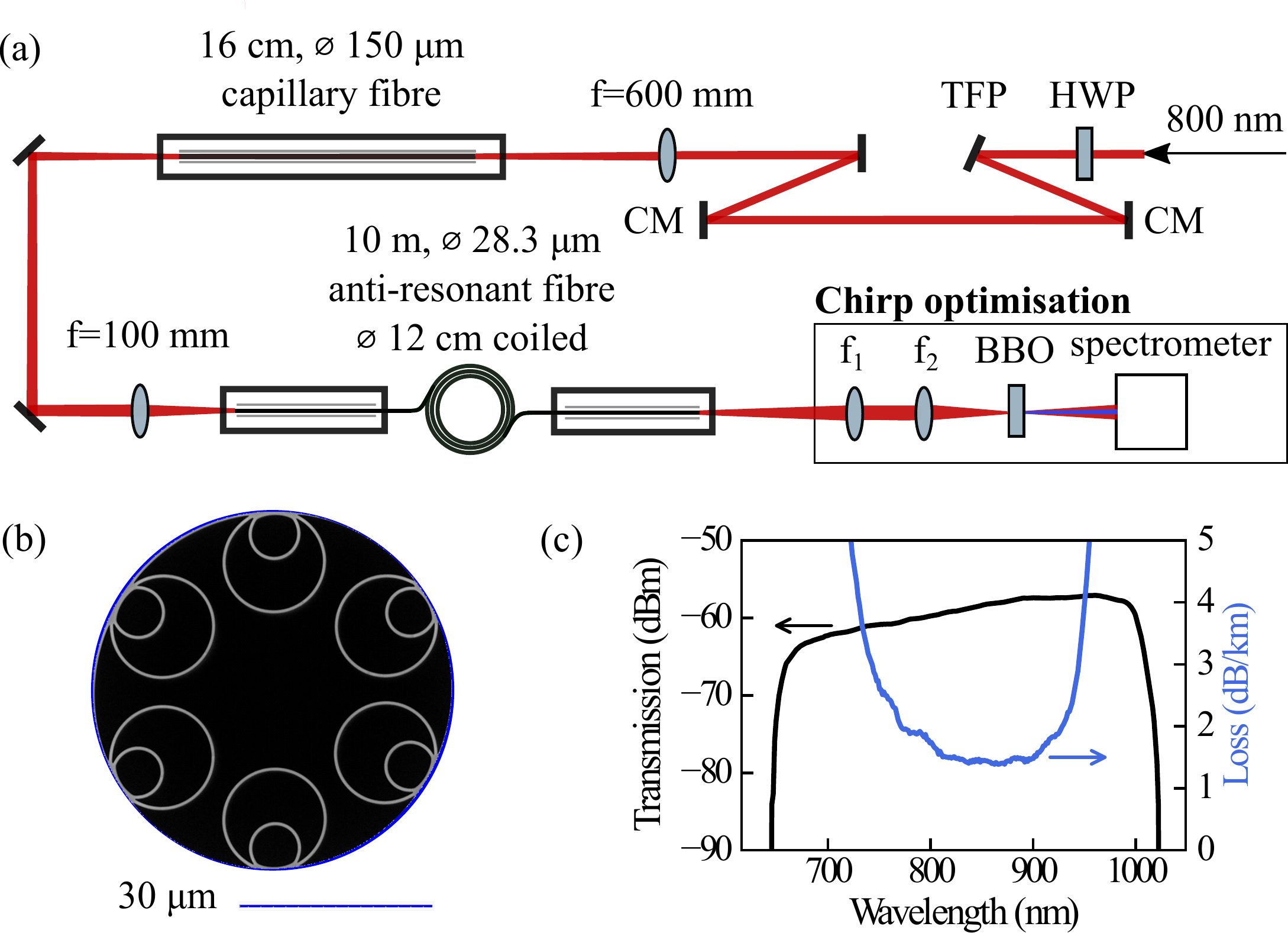}
\caption{(a) Sketch of the experimental setup. HWP: half-wave plate, TFP: thin-film polarizer, CM: chirped mirror, f$_1$: $100$~mm collimating lens, f$_2$: $150$~mm focusing lens. (b) fiber cross section with \SI{28.3}{\micro\meter} core diameter. (c)~fiber transmission (black) as measured for 10~m fiber length, and attenuation (blue)~\cite{sakr_hollow_2020}.}
\label{fig:fig1}
\end{figure}

The primary experimental setup is shown in Fig.~\ref{fig:fig1}(a). A Ti:Sapphire laser provided pulses at 800~nm with 1~kHz repetition rate. A thin-film polarizer and a half-wave plate were used for power control and two chirped mirrors (-250~fs$^2$ each, Ultrafast Innovations) were used for dispersion control. A 16~cm long, \SI{150}{\micro\meter} core diameter stretched hollow capillary fiber~\cite{nagy_flexible_2008,travers_high-energy_2019} was kept in low vacuum and used as a spatial filter to optimize the beam quality and stabilize the laser pointing. The transmission through the capillary fiber was 70\%, corresponding to a coupling efficiency of 84\%, including correction for the Fresnel reflections from the windows. The pointing fluctuations were 4 times reduced compared to the beam entering the capillary.

A single achromatic lens (f=100~mm) was used to focus the beam to a \SI{21}{\micro\meter} diameter spot ($1/\mathrm{e}^2$) and couple the laser pulses into a 10~m long, \SI{28.3}{\micro\meter} core diameter nested anti-resonant fiber~\cite{sakr_hollow_2020}. The optimal spot size at the anti-resonant fiber entrance was determined by varying the position of the lens and fiber input facet, using the divergence of the beam exiting the capillary. At the fiber output, the pulses were characterised by a spectrometer and a home-built second harmonic generation frequency-resolved optical gating (SHG-FROG) setup. To achieve compressed pulses at the fiber exit, we optimized the chirp of the initial laser pulse. The beam exiting the fiber was collimated (f$_1$=100~mm, N-BK7, 3.6~mm thick) and then focused (f$_2$=150~mm, fused silica, 3.2~mm thick) onto a \qty{10}{\um} thick $\beta$-barium borate (BBO) crystal. The grating compressor of the laser system was then tuned to maximise the second-harmonic signal.

Fig.~\ref{fig:fig1}(b) shows a scanning electron micrograph of the anti-resonant fiber cross section, with a calculated mode field diameter of \SI{19.4}{\micro\meter} and wall thickness of \SI{585}{\nano\meter}. The group-velocity dispersion and third-order dispersion of the fiber were calculated using the capillary model\cite{marcatili_hollow_1964} to be -663~fs$^2$/m and 863~fs$^3$/m at 800~nm, respectively. The fiber has a large transmission band around 800~nm, where it exhibits ultra-low loss down to 1.5~dB/km, as measured by a cutback from 1060~m to 10~m length with a white light source~\cite{sakr_hollow_2020} and shown in Fig.~\ref{fig:fig1}(d). The 10~m long fiber was tightly coiled with 6~cm radius and its two ends were mounted into cells that could be sealed and evacuated independently. Optical access was provided by two 1~mm thick uncoated MgF$_2$ windows. The fiber was evacuated and the vacuum level was stabilised at 1.5~$\times~10^{-2}$~mbar by continuously operating a roughing pump. 

\section{Plasma and vacuum considerations}

Even under low vacuum conditions, plasma dynamics can become significant when high pulse intensities are present~\cite{saleh_understanding_2011}. To determine the role of plasma in our experiment, we used numerical simulations~\cite{travers_high-energy_2019, brahms_lunajl_2021} where the dispersion of the hollow fiber was approximated using the Marcatili capillary model~\cite{marcatili_hollow_1964} without linear attenuation. We modelled chirped input pulses (35~fs bandwidth-limited Gaussian pulses positively chirped to 500~fs) propagating through the fiber filled with \SI{1.5e-2}{\milli\bar} of oxygen (O$_2$) to simulate the remnant gas density inside the fiber after rough pumping. Plasma dynamics were modelled using the model detailed in Ref.~\cite{geissler_light_1999} with the ionisation rate calculated with the PPT model~\cite{perelomov_ionization_1966, talebpour_semi-empirical_1999}.

\begin{figure}[ht]
\centering
\includegraphics[width=1\linewidth]{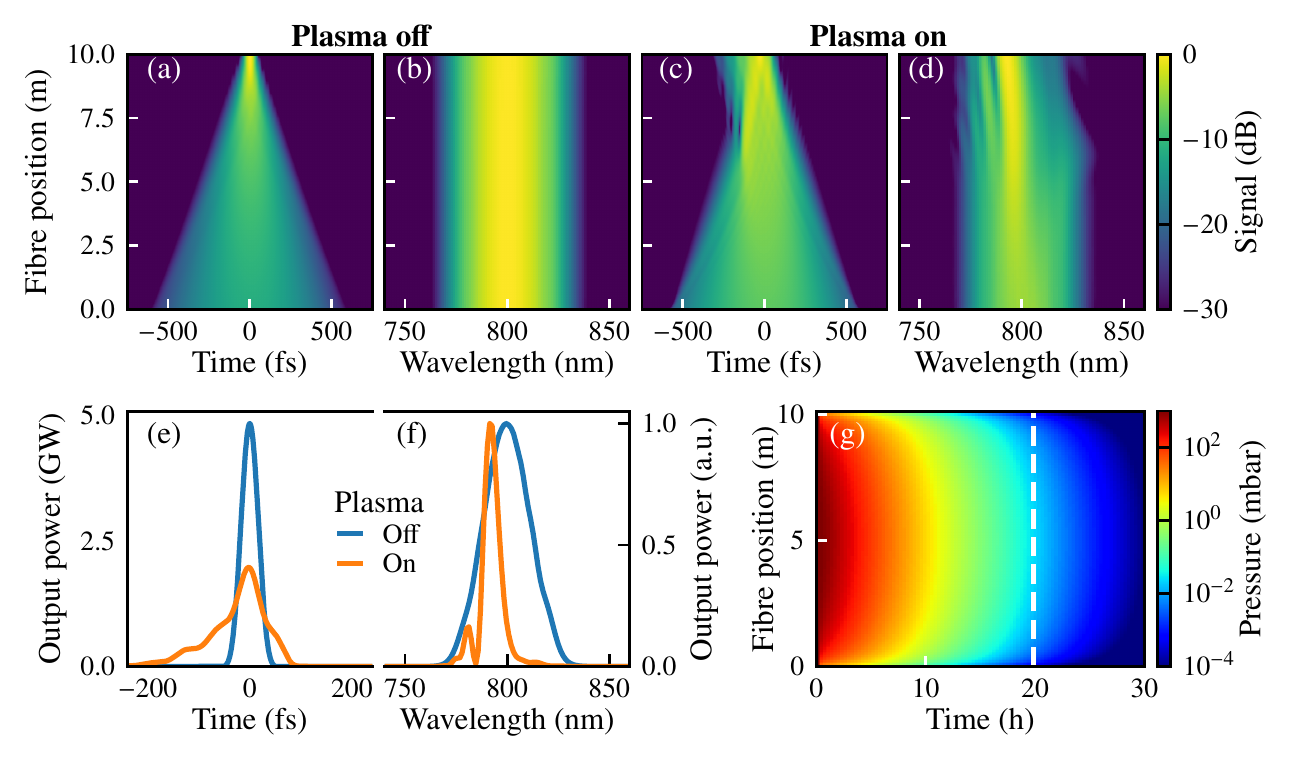}
\caption{Simulated propagation of an 800~nm, 35~fs bandwidth-limited Gaussian pulse positively chirped to 500~fs at \qty{200}{\micro\joule} energy and \SI{1.5e-2}{\milli\bar} oxygen pressure with plasma off (a, b) and plasma on (c, d), respectively. The ionisation rate is calculated using the PPT model~\cite{perelomov_ionization_1966}. Pulse profiles (e) and power spectra (f) at the fiber output for the cases above. (g) Estimated evacuation evolution along the fiber position with time. The dashed white line annotates the time required to reach \SI{1.5e-2}{\milli\bar} at the fiber mid-point.}
\label{fig:fig2}
\end{figure}

Fig.~\ref{fig:fig2} shows the results for \qty{200}{\micro\joule} pulse energy. The temporal and spectral evolution along the fiber length is shown for two modelling conditions: (i) when the plasma dynamics are turned off and (ii) when the plasma dynamics are turned on. In the former case, the pulse smoothly compresses to its bandwidth-limited shape along the fiber, because the initial positive chirp is compensated by the negative dispersion of the fiber, and its spectral profile is delivered to the fiber output unchanged. In the latter case, the gas is fully ionised along the full length of the fiber, and both the temporal profile and the spectrum of the pulse are distorted.

This distortion becomes more evident in  Fig.~\ref{fig:fig2}(e,f), where the pulse profiles and power spectra at the fiber output are plotted for both cases. The presence of plasma causes a severe drop in power and stretches the pulse in the time domain, while blue-shifting and spectral narrowing is induced in the spectral domain. This is due to the high intensity, which is sufficient to fully ionise the residual gas inside the fiber. Because the gas is fully ionised, the resulting plasma effects on the pulse propagation are significant despite the low pressure. In practice, this means that for efficient delivery of high-intensity pulses in long fibers, even higher vacuum is required (less than \SI{e-3}{\milli\bar} for our parameters) and that this vacuum must be achieved throughout the fiber length. 

In Fig.~\ref{fig:fig2}(g) we consider a model of gas flow dynamics in hollow core fibers, as described in Ref.~\cite{henningsen_dynamics_2008}, and show an estimated pressure evolution with time along the fiber when evacuation is initiated from its two end-points. Due to the fiber's small diameter and long length, it takes up to 20~hours for the pressure at the mid-point position to match the pressure at the end-points, which for our experiment was stabilised and measured at \SI{1.5e-2}{\milli\bar} [dashed line in Fig.~\ref{fig:fig2}(g)]. To ensure adequate evacuation throughout its total length, the fiber was purged with helium and subsequently evacuated with a roughing pump from both ends for at least 24 hours.

\section{Input pulses and energy transmission}\label{sec:ET}

\begin{figure}[ht]
\centering
\includegraphics[width=1\linewidth]{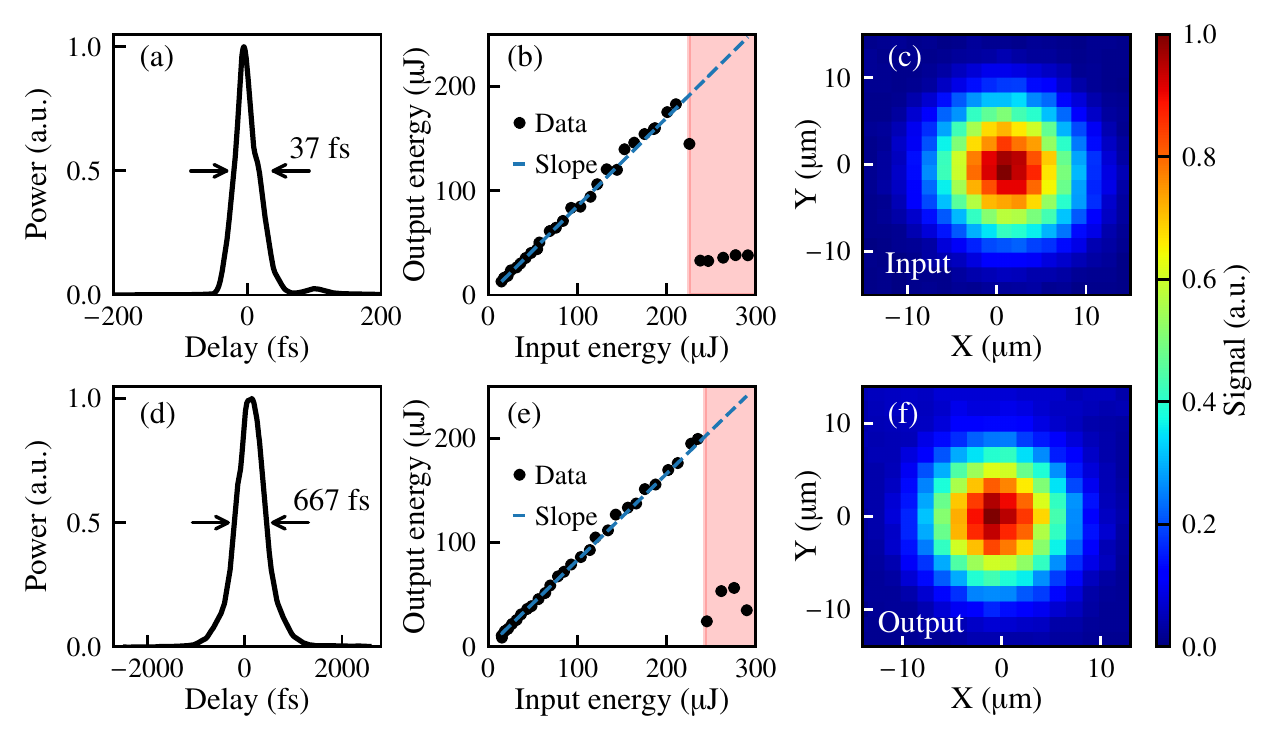}
\caption{Input pulse (a) and energy transmission (b) when pumping with compressed input pulses. Transmission efficiency (slope) is 85\%. Auto-correlation trace (d) and energy transmission (e) when pumping with stretched input pulses. The estimated pulse duration is 470~fs assuming a Gaussian pulse shape. Transmission efficiency (slope) is 83\%. The shaded areas in (b) and (e) annotate the energy range where guidance was lost. Near-field beam profiles at fiber input (c) and output (f).}
\label{fig:fig3}
\end{figure}

\begin{figure}[ht]
\centering
\includegraphics[width=0.60\linewidth]{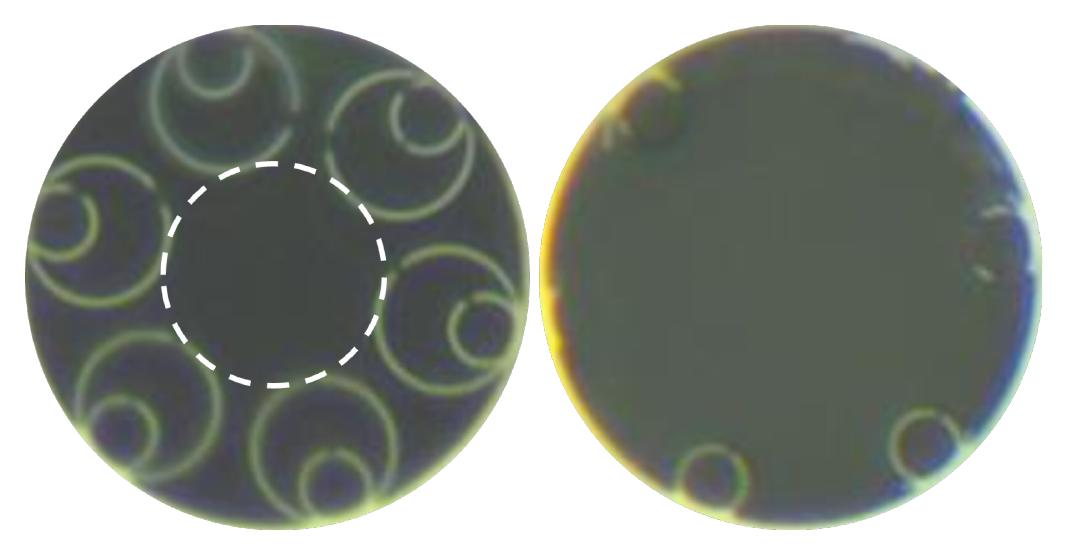}
\caption{Microscope images of the input facet of the anti-resonant fiber before and after damage. The white dashed circle annotates the core-wall interface where optical damage is initiated.}
\label{fig:fig10}
\end{figure}

We initially pumped the evacuated fiber with the compressed laser pulses without using the chirp optimization at the fiber output. The duration of the compressed pulses was 37~fs full width at half maximum (FWHM), as measured with our home-built SHG-FROG and shown in Fig.~\ref{fig:fig3}(a). Pulse energies up to \qty{210}{\micro\joule} were successfully coupled into the fiber as shown in Fig.~\ref{fig:fig3}(b). This corresponds to a very high input peak intensity of \qty{3}{\peta\watt\per\square\cm} and a fluence at the inner interface of the core-wall of \qty{3}{\joule\per\square\cm}, which is close to the damage threshold of silica glass at 40~fs pulse duration~\cite{chimier_damage_2011, hoffart_surface_2011}. For fluence levels beyond this threshold we observed loss of guidance and damage to the fiber entrance facet [see shaded area in Fig.~\ref{fig:fig3}(b)]. Microscope images of the fiber input facet before and after damage are shown in Fig.~\ref{fig:fig10}. The inner core-wall interface is highlighted with a white dashed circle. After damage, the inner structure of the fiber is almost completely lost, including some of the nested tubes. Further results on the anti-resonant fiber fluence limits are presented in section~\ref{sec:mJenergy}. Here, the average transmission efficiency was 85\%, including correction for the input windows. Subsequent cut-back measurements at \SI{200}{\micro\joule} input energy showed 90\% transmission in a 2~cm long piece, meaning that only a small fraction of the energy (5\%) was lost due to fiber attenuation, despite the tight fiber coiling.

In order to deliver compressed pulses to a target at the output of the fiber, we tuned the grating compressor of the laser to chirp the input pulse. With the stretched input pulses, pulse energies close to \SI{250}{\micro\joule} were coupled into the fiber, with an 83\% average transmission efficiency and maximum output pulse energy of \SI{200}{\micro\joule}, as shown in Fig.~\ref{fig:fig3}(e). The higher input fluence achieved in this case is in accordance with the increase of the glass damage threshold when pumping with longer pulses~\cite{chimier_damage_2011}. A second-order autocorrelation trace of the stretched input pulse is shown in Fig.~\ref{fig:fig3}(d). The width of 667~fs corresponds to an estimated pulse duration of 470~fs, assuming a Gaussian pulse shape. With both compressed and stretched input pulses, the beam exited the fiber in the fundamental mode. The near-field spatial profile after re-imaging to match the size of the near-field profile at the fiber input [see Fig.~\ref{fig:fig3}(c)] is shown in Fig.~\ref{fig:fig3}(f). The beam profiles were captured with the highest resolution camera available, with \SI{1.67}{\micro\meter} pixel size.

\section{Spectrum relaxation and plasma suppression}
\begin{figure}[ht]
\centering
\includegraphics[width=1\linewidth]{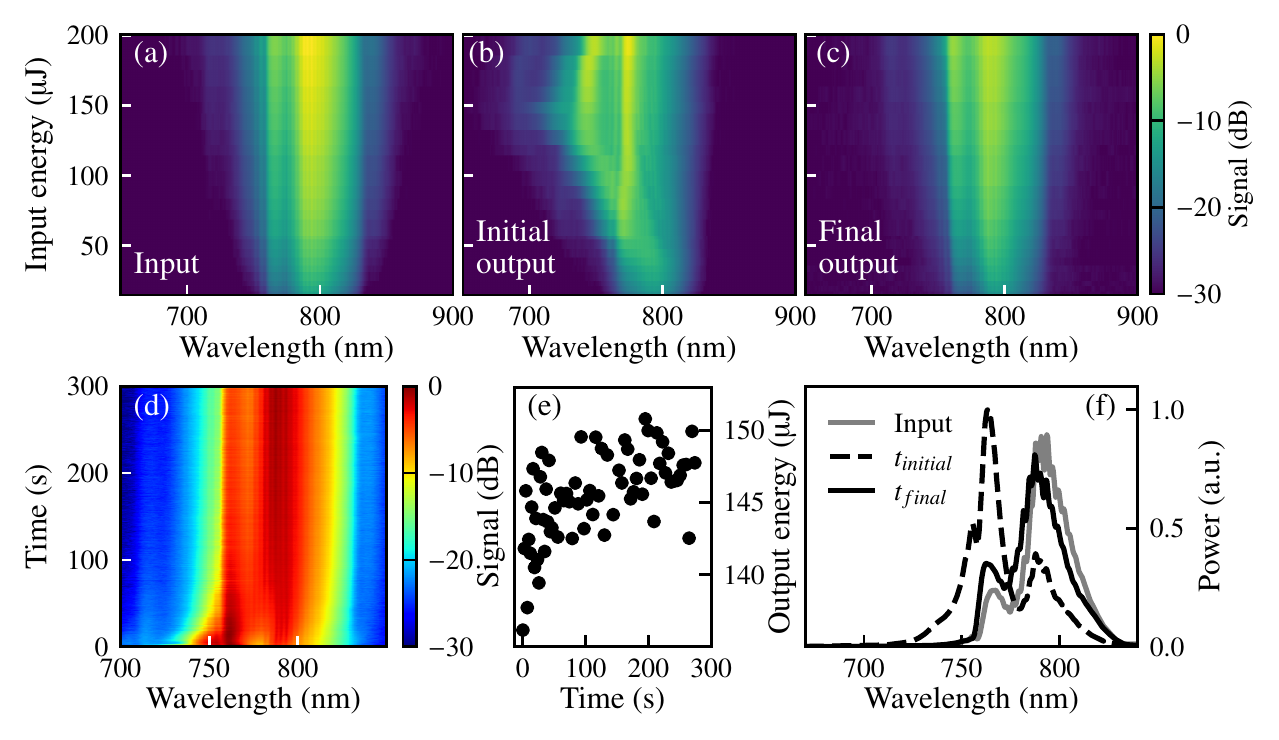}
\caption{Spectra with increasing pulse energy at the fiber input (a) and at the output (b) without pre-pumping the fiber with the laser pulses. (c) Output spectra after pre-pumping the fiber with the laser pulses. (d) Evolution of the output spectrum with time when continuously pumping the fiber with \qty{180}{\micro\joule} pulse energy. (e) Output energy while pumping the fiber with laser pulses. (f) Comparison between input spectrum, and initial and final output spectra at \qty{180}{\micro\joule} energy.}
\label{fig:fig4}
\end{figure}

Using the chirped input pulses and pulse energies up to about \SI{200}{\micro\joule}, we investigated the spectrum of the pulses delivered through the evacuated fiber. In Fig.~\ref{fig:fig4}(a) and (b), the input and output spectra are plotted against the input pulse energy on a logarithmic color scale. The output spectra exhibited large spectral broadening down to 700~nm, despite the systematic fiber evacuation. This is expected from the very high pulse intensity, and the simulations shown in Fig.~\ref{fig:fig2}. We later observed that the spectral components were slowly red-shifting back to the wavelengths of the input spectrum over the course of several minutes.

This evolution of the output spectra with time was captured at \SI{180}{\micro\joule} input energy and it is shown in Fig.~\ref{fig:fig4}(d). Within the first few-tens seconds the spectrum almost entirely shifts back to its input shape and gradually re-builds its peak at around 800~nm within the next few minutes. After about 5 minutes, the output spectrum closely resembles the input. This process results in higher fiber transmission, as shown in Fig.~\ref{fig:fig4}(e), because the pulse spectrum is more closely confined in the low-loss band of the fiber. The input spectra together with the initial and final output spectra at \SI{180}{\micro\joule} input energy are shown in Fig.~\ref{fig:fig4}(f). The resulting output spectrum (solid black) reverted almost entirely back to its input shape (solid gray) with only a slight blue-shift. We then repeated the measurement of the output spectrum as a function of input energy and observed negligible spectral broadening at all input energies, as shown in Fig.~\ref{fig:fig4}(c). For both Fig.~\ref{fig:fig4}(b) and Fig~\ref{fig:fig4}(c) individual spectra were obtained within a few-tens of milliseconds, and the total scan duration was of the order of a few seconds. In the following discussion we refer to this reverting of the output spectrum back to a spectrum resembling the input as ``spectrum relaxation''.

We have established that once the fiber is pre-pumped with laser pulses, the relaxed spectrum output is constantly reproduced, as long as the vacuum level remains intact, even when the pump laser is blocked. Instead, the non-relaxed, broadened, output of Fig.~\ref{fig:fig4}(b) only re-appears once the vacuum breaks, air is allowed to enter the fiber, and the fiber is evacuated again. This was tested by pre-pumping the fiber to achieve spectrum relaxation and allowing for 72 hours, during which the pump beam was blocked and the vacuum remained intact. After 72 hours, we re-pumped the fiber with laser pulses and observed the same relaxed spectrum output as in Fig.~\ref{fig:fig4}(c). To reset the relaxed spectrum output, we had to break the vacuum, allowing air to enter the fiber. After subsequent 24-hour evacuation, the broadened output of Fig.~\ref{fig:fig4}(b) re-appeared. We have also noticed that the magnitude of the spectral relaxation is proportional to the input energy, with negligible dynamics observed when pre-pumping at lower energies ($<\SI{20}{\micro\joule}$).


These results were surprising to us. As a result of the very high intensity present inside the fiber core ($\sim\text{PW/}\text{cm}^2$), any gas molecules near the centre of the core will be ionised. After the removal of the pump light, the electrons and ions would be expected to recombine and relax back to a neutral gas. If this was the case, ionisation would restart and plasma would be re-created once the pump light was switched on again, which does not occur in our experiment. The suppression of traces of plasma effects on the output spectrum over long periods of time suggests that the molecules have instead been removed from the centre of the fiber core. It appears that the fiber pre-pumping acts as an active ``gas suppression'' mechanism, mitigating the effects of photo-ionisation and allowing for effective delivery of intense pulses through the fiber, similar to the plasma-free propagation of Fig.~\ref{fig:fig2}.

One hypothesis to explain this is that the ionised remnant air molecules within the core become attached to the inner glass capillary surfaces, in a process similar to surface recombination~\cite{kim_recombination_1991, marinov_adsorption_2014}. This removal of molecules from the hollow core remains semi-permanent and only resets once the fiber is re-filled with air. This hypothesis is further supported by considering the inter-molecular mean free path (average path between molecular collisions), which at this pressure is more than two orders of magnitude larger than the fiber core. Therefore, the number of ion collisions with the glass core-wall is far greater than collisions between molecules, ions and electrons. However, it must be stressed that detailed further investigation is required to understand the origin of these observations.

\section{Output pulse profiles}

\begin{figure*}[ht]
\centering
\includegraphics[width=1\linewidth]{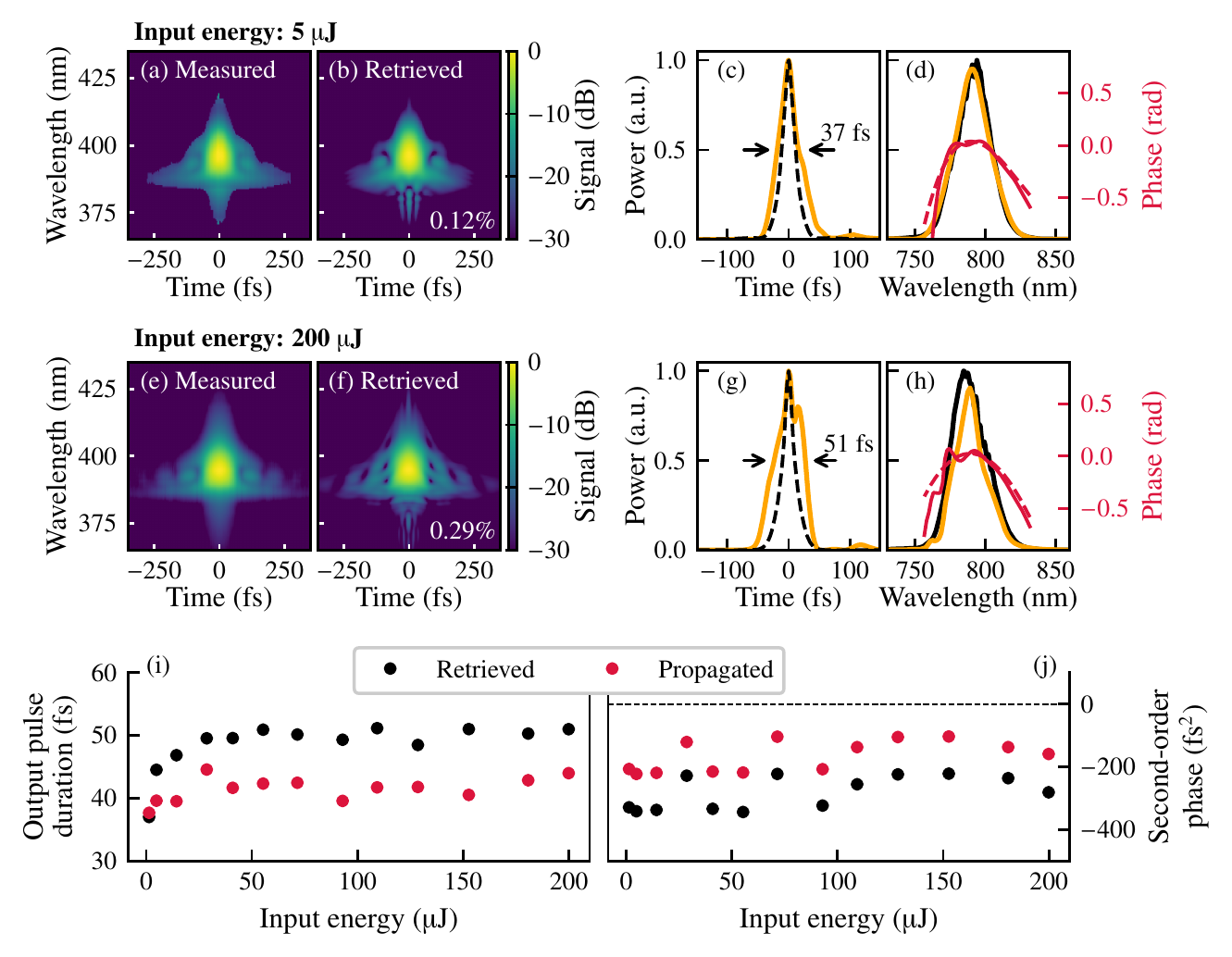}
\caption{SHG-FROG measurements of the output pulses at \qty{5}{\micro\joule} (a-d) and \qty{200}{\micro\joule} (e-h) input energy, with measured (a, e) and reconstructed (b, f) traces in logarithmic color scale. (c, g) Retrieved (orange) and bandwidth-limited pulses (dashed black). (d, h) Retrieved spectral profiles (orange) and phases (solid red), fitted phases (dashed red) and externally measured power spectra (solid black). Evolution of output pulse duration (i) and second-order phase (j) with input energy for the measured pulses (black dots) and for pulses numerically propagated (red dots) through 3.2~mm of fused silica.}
\label{fig:fig5}
\end{figure*}

The output pulses delivered after the spectrum relaxation were measured as a function of input energy using the SHG-FROG after the collimating lens (f$_1$) of the chirp optimization layout. Fig.~\ref{fig:fig5}(a-h) shows the measured and reconstructed traces and the retrieved pulses at low (\SI{5}{\micro\joule}) and high (\SI{200}{\micro\joule}) input energy. The FROG error is low (<0.3\%) and the agreement between the retrieved and the externally measured spectrum is excellent in both cases. From an input pulse duration of approximately 470~fs, the delivered pulse duration was 35~fs at low energy and 51~fs at high energy. The retrieved bandwidth-limited pulses were approximately 25~fs long. 

The evolution of the retrieved pulse duration and second-order phase with energy is plotted with black dots in Fig.~\ref{fig:fig5}(i,j). The retrieved output pulses were negatively chirped and constant at about 50~fs duration for pulse energies above \SI{25}{\micro\joule}. The chirp optimisation was deliberately performed at low energies (<\SI{5}{\micro\joule}) to assess the effect of the plasma-suppression on the output pulses. The clamping on the measured pulse duration for energies above the onset of spectrum relaxation ($\sim\SI{20}{\micro\joule}$) suggests that the plasma-induced pulse stretching was largely suppressed. This is also evident in the evolution of the second-order phase, where no additional plasma-induced anomalous dispersion was detected at higher energies. The sign of the retrieved spectral phases was cross-checked by measuring the FROG traces with and without additional glass in the beam path.

The negative chirp of the output pulses can be easily compensated, as shown with red dots in Fig.~\ref{fig:fig5}(i,j), so that the delivered on-target pulses are compressed to 40~fs, corresponding to a maximum delivered peak power of about 5~GW. These results were obtained by numerically propagating the retrieved pulses through a 3.2~mm thick fused silica lens, the same as the focusing lens that was used in chirp optimisation (f$_2$ in Fig.~\ref{fig:fig1}), which was removed for the FROG measurements. We believe that further compression towards the bandwidth-limited duration was probably limited by the chirped mirrors and the fiber-induced third-order dispersion.

\section{Energy scaling to mJ-level}\label{sec:mJenergy}
As shown in section~\ref{sec:ET}, pulse energies up to \SI{210}{\micro\joule} were efficiently coupled into the \SI{28.3}{\micro\meter} core diameter anti-resonant fiber when using compressed input pulses. For higher input energies, we observed loss of guidance and damage at the input facet of the fiber. The damage occurred when the fluence at the inner interface of the core-wall approached the damage threshold of silica glass, which is around \qty{3}{\joule\per\cm\squared} at 40~fs pulse duration~\cite{hoffart_surface_2011}. For a Gaussian beam profile at the input facet of the fiber, the fluence is given by:
\begin{equation}\label{eqn:fluence}
    F(E, r, w)=\frac{2E}{\pi w^2}e^{{-2}{\frac{r^2}{w^2}}},
\end{equation}
where $E$ is the pulse energy, $w$ is the $1/\mathrm{e}^2$ beam radius at the input facet of the fiber and $r$ is radial distance. To approximate the fluence at the core wall, we set the $r$ parameter equal to the fiber core radius, so that $r=\qty{14.15}{\micro\meter}$ and $w=\qty{10.5(0.5)}{\micro\meter}$, which results in a fluence at the core-wall of \qty{3.2(0.8)}{\joule\per\square\cm}. The Gaussian approximation in Eq.~\ref{eqn:fluence} is valid for the incident beam at the input facet of the fiber. Inside the fiber, this approximation is less accurate, especially close to the silica membranes which form the core wall. Nevertheless, finite-element simulations confirm that for the guided mode inside the fiber, the fluence at the boundary of the silica membranes is similar to the Gaussian approximation, suggesting that the damage threshold at the input facet is not substantially lower than that inside the fiber itself. This suggests that even if the mode-matching of the incident beam were improved beyond what is possible with a Gaussian beam profile, this would only marginally increase the damage threshold of the fiber.

\begin{figure*}[ht]
\centering
\includegraphics[width=1\linewidth]{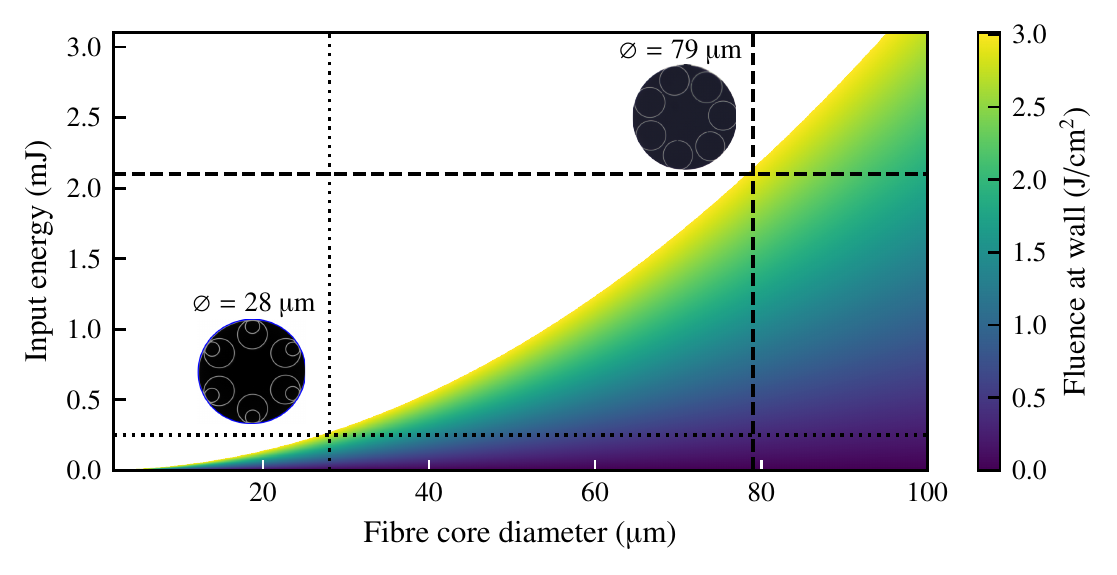}
\caption{Below-threshold fluence (up to \qty{3}{\joule\per\cm\squared} for 40~fs long pulses) at the fiber core-wall with respect to input energy and fiber core diameter, as calculated using Eq.~\ref{eqn:fluence}. The dotted lines annotate the maximum energy predicted for a \SI{28}{\micro\meter} diameter fiber and the dashed lines annotate the maximum energy predicted for a \SI{79}{\micro\meter} diameter fiber.}
\label{fig:fig7}
\end{figure*}

\begin{figure*}[ht]
\centering
\includegraphics[width=1\linewidth]{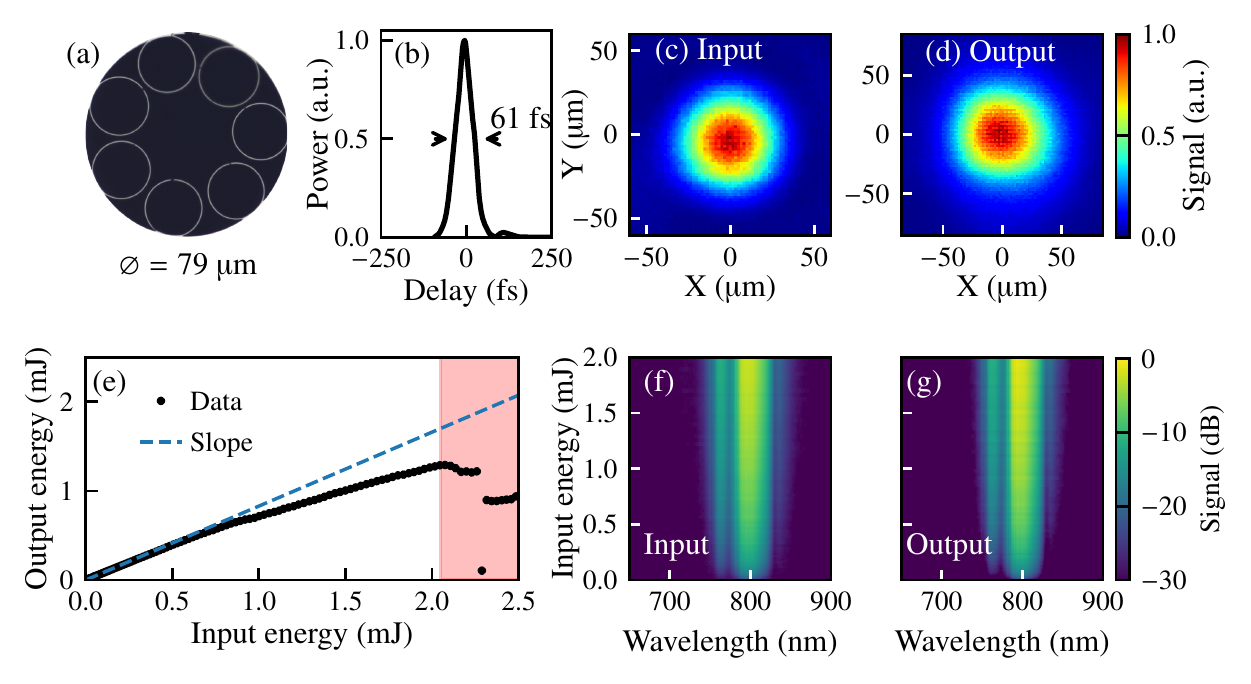}
\caption{(a) Cross-section of the larger-core (\SI{79}{\micro\meter} diameter) anti-resonant fiber used to scale the input energy. (b) Input pulse and input (c) and output (d) near-field beam profiles. (e) Energy transmission of large-core fiber, with slope efficiency (dashed blue) of 85\%. The shaded area annotates the energy regime where guidance is lost. Input (f) and output (g) spectra with increasing energy.}
\label{fig:fig8}
\end{figure*}

From the above equation it is evident that scaling the input pulse energy can be achieved by increasing the fiber core size and hence the spot size $w$, so that the fluence at the fiber core-wall remains below the damage threshold. This scaling mechanism allows mJ-level pulse delivery to be performed using anti-resonant hollow fibers, as shown in Fig.~\ref{fig:fig7}. Here, we show the below-threshold fluence (up to \qty{3}{\joule\per\cm\squared} for 40~fs long pulses) at the core-wall for pulse energies up to \SI{3}{\milli\joule} and fiber core diameters up to \SI{100}{\micro\meter}, assuming Gaussian input beam radii of $w=0.7r$. These calculations predict that a maximum coupled energy of \SI{0.25}{\milli\joule} can be achieved using a \SI{28.3}{\micro\meter} fiber core diameter [see dotted lines in Fig.~\ref{fig:fig7}], which is close to the maximum coupled energy achieved experimentally. For mJ-level transmission, fiber core diameters above \SI{60}{\micro\meter} are required, while at \SI{79}{\micro\meter} diameter the maximum coupled energy can reach \SI{2.1}{\milli\joule} [see dashed lines in Fig.~\ref{fig:fig7}].

To further confirm our calculations, we tested the energy limits of a few-cm long, large-core (\SI{79}{\micro\meter} diameter) hollow-core anti-resonant fiber, the cross-section of which is shown in fig.~\ref{fig:fig8}(a). Although the increased attenuation of this fiber at 800~nm did not allow for laser delivery over several meters, as achieved with the smaller core fiber, it was used to determine the maximum fluence at the input core-wall interface before damage. For this experiment we picked off the 800~nm pulses after the HWP and TFP shown in Fig.~\ref{fig:fig1}, bypassing the chirped mirrors and the capillary filter. As a result, the input pulses were slightly stretched to 61~fs, as measured with the SHG-FROG and shown in Fig.~\ref{fig:fig8}(b). The fiber was housed in a single cell and was evacuated. Optical access was provided by 1~mm thick fused silica windows. A single lens (f=300~mm) was used to focus the beam to a $w=\qty{28.5(0.5)}{\um}$ spot. The spatial profile of the focus is shown in Fig.~\ref{fig:fig8}(c). Coupling into the fiber was optimised at low energy and a near-field output is shown in Fig.~\ref{fig:fig8}(d).

As predicted in Fig.~\ref{fig:fig7}, pulse energies up to \SI{2.1}{\milli\joule} were successfully coupled into the fiber, with a transmission efficiency of 83\% achieved for pulse energies up to \SI{0.8}{\milli\joule} [see dashed line in Fig.~\ref{fig:fig8}(e)]. For input energies between 0.8 and \SI{2.1}{\milli\joule}, fiber transmission deteriorated to 65\%, possibly due to nonlinear lensing on the input window. For input energies above \SI{2.1}{\milli\joule} [shaded area in Fig.~\ref{fig:fig8}(e)] energy transmission drops and loss of guidance occurs. The input and output spectra [shown in Fig.~\ref{fig:fig8}(f) and Fig.~\ref{fig:fig8}(g), respectively] are very similar, suggesting plasma-free laser propagation. The maximum guided peak intensity was 2.5~PW/cm$^2$ and the maximum fluence at the input core-wall interface was \qty{3.5(0.3)}{\joule\per\square\cm}, similar to the fluence calculated for the smaller core fiber. At the fiber output, the maximum energy was \SI{1.3}{\milli\joule}, corresponding to \SI{20}{\giga\watt} of peak power.

Our results are consistent with previous pulse delivery experiments, where even higher pulse energies (\SI{2.6}{\milli\joule}) were delivered over fibers with larger core diameter~(\SI{100}{\micro\meter})~\cite{debord_26_2015}. Despite the higher energy, the in-core intensity was similar and the maximum fluence at the core-wall is estimated to be about $3\ \text{J/}\text{cm}^2$, very close to the values achieved here. These results confirm that even at the presence of high-intensity fields in the core, the fluence at the core-wall interface is the natural damage limit to high energy transmission through these types of fibers.

\section*{Conclusion}
We have demonstrated efficient on-target delivery of 40~fs, 5~GW pulses through an evacuated 10~m long, 6~cm coil radius, hollow nested anti-resonant fiber with small core from chirped 470~fs long laser pulses at 800~nm. With compressed input pulses, we achieved a guided peak intensity of \qty{3}{\peta\watt\per\square\cm} and a fluence at the inner interface of the core-wall of \qty{3}{\joule\per\square\cm}. We observed that when initially pumping the fiber with intense pulses, after evacuation, the output pulse spectrum shows initial signs of plasma blue-shift but then relaxes back to the input spectrum after a few minutes. It then consistently remains like this so long as the low vacuum is maintained, even after blocking the pump pulses, and over time-scales of at least several days. This remarkable observation suggests a mechanism of gas suppression or removal from the hollow core, which we attribute to adsorption of the ionised particles at the silica surfaces.

This effect allows for the scaling of the pulse energy to the mJ-level by using larger fiber core diameters to increase the energy transmission while keeping the laser fluence on the core-wall below the damage threshold of the glass. We have experimentally verified the energy scaling by successfully coupling \SI{2.1}{\milli\joule} into a larger core fiber in accordance with the fluence-imposed limits of the glass, to obtain 20~GW of peak power at the output of the fiber. Our results can pave the way towards the integration of anti-resonant hollow fibers for flexible delivery of high intensity pulses on target in nonlinear post-compression schemes and in laser-source development.

\section*{Acknowledgements}
This work was funded by the EPSRC (Airguide Photonics EP/P030181/1) and the ERC (LightPipe, g.a. 682724; and XSOL, g.a. 101001534). F.B.~acknowledges support from the Royal Academy of Engineering through Research Fellowship No.~RF/202021/20/310. C.B.~acknowledges support from the Royal Academy of Engineering through Research Fellowship No.~RF/202122/21/133.

\section*{Disclosures}
The authors declare no conflicts of interest. 

\bibliography{MyLibrary}

\end{document}